\documentclass[11pt]{article}

\usepackage{graphicx}    % standard LaTeX graphics tool
\usepackage{amsfonts}
\usepackage{bm}

\def\beq#1{{\[#1\]}}
\def\be#1\ee{\begin{equation}#1\end{equation}}
\def\ba#1#2\ea{\begin{eqnarray}#1#2\end{eqnarray}}
\newtheorem{remark}{Remark}[section]
\def\gt{\tilde g}
\def\ft{\tilde f}
\def\St{\tilde S}
\def\ff{\hat f}
\def\mut{\tilde \mu}
\def\bigM{ {\mathcal {M}}}
\def\bigF{ {\mathcal {F}}}
\def\real{ {\mathbb {R}}}
\newcommand{\bq}{\begin{equation}}
\newcommand{\eq}{\end{equation}}
\newtheorem{theorem}{Theorem}[section]

\begin{document}
\title{Mesoscopic modelling of financial markets}

\author{Stephane Cordier\thanks{MAPMO Universit\'e d'Orl\'eans and CNRS, (UMR 6628)
 F\'ed\'eration Denis Poisson, (FR 2964)
 BP 6759, 45067 Orl\'eans cedex 2, France.
(\texttt{stephane.cordier@math.cnrs.fr})}\,\and Lorenzo
Pareschi\thanks{Department of Mathematics and CMCS, University of
Ferrara, Via Machiavelli 35 I-44100 Ferrara, Italy.
(\texttt{lorenzo.pareschi@unife.it})}\,
 \and Cyrille Piatecki\thanks{Laboratoire d'Economie d'Orléans (LEO) UMR 6221
University of Orl\'eans and CNRS, 45067 Orl\'eans, France.
(\texttt{cyrille.piatecki@univ-orleans.fr})}}

\date{October 24, 2008}

\maketitle

\begin{abstract}
We derive a mesoscopic description of the behavior of a simple
financial market where the agents can create their own portfolio
between two investment alternatives: a stock and a bond. The
model is derived starting from the Levy-Levy-Solomon microscopic
model \cite{LLS, LLS1} using the methods of kinetic theory and
consists of a linear Boltzmann equation for the wealth
distribution of the agents coupled with an equation for the price
of the stock. From this model, under a suitable scaling, we derive a
Fokker-Planck equation and show that the equation admits a
self-similar lognormal behavior. Several numerical examples are
also reported to validate our analysis.
\end{abstract}

\noindent {\bf Keywords:} wealth distribution, power-law tails,
stock market, self-similarity, kinetic equations.

%\tableofcontents

\section{Introduction}
In recent years, physicists have been growing more and more
interested in new interdisciplinary areas such as sociology and
economics, originating what is today named socio-economical
physics \cite{BM, AC1, CCM, YD, GGPS, IKR, LLS, MS, PGS, VT}. This
new area in physics borrows several methods and tools from
classical statistical mechanics, where complex behavior arises
from relatively simple rules due to the interaction of a large
number of components. The motivation behind this is the attempt to
identify and characterize universal and non-universal features in
economical data in general.

A large part of the research in this area is concerned with
power-law tails with universal exponents, as was predicted more
than one century ago by Pareto \cite{CCM, LLS2, Pa}. In particular,
by identifying the wealth in an economic system with the energy of
a physical system, the application of statistical physics makes
it possible to understand better the development of tails in
wealth distributions. Starting from the microscopic dynamics,
mesoscopic models can be derived with the tools of classical
kinetic theory of fluids \cite{BM, CPT, YD, DT, IKR, MT, PT1, SL}.

In contrast with microscopic dynamics, where behavior often
can be studied only empirically through computer simulations,
kinetic models based on PDEs allow us to derive analytically general
information on the model and its asymptotic behavior. For example,
the knowledge of the large-wealth behavior is of primary
importance, since it determines a posteriori whether the model can fit
data of real economies.

In some recent papers, the explicit emergence of power laws in the
wealth distribution, with Pareto index strictly larger than one,
has been proved for open market economies where agents can
interact through binary exchanges together with a simple source of
speculative trading \cite{BM, CPT, PT1, SL}.

The present work is motivated by the necessity to have a more
realistic description of the speculative dynamics in the above
models. To this end, we derive a mesoscopic description of the
behavior of a simple financial market where a population of
homogeneous agents can create their own portfolio between two
investment alternatives: a stock and a bond. The model is closely
related to the Levy-Levy-Solomon (LLS) microscopic model in
finance \cite{LLS, LLS1}. This model attempted to construct from
simple rules complex behavior that could then mimic the market and
explain the price formation mechanism. As a first step towards a
more realistic description, we derive and analyze the model in the
case of a single stock and under the assumption that the optimal
proportion of investments is a function of the price only. In
principle, several generalizations are possible (different stocks,
heterogeneous agents, a time-dependent optimal proportion of
investments, \ldots), and we leave them for future investigations.

In our non-stationary financial market model, the average wealth is
not conserved and this produces price variations. Let us point out
that, even if the model is linear since no binary interaction
dynamic between agents is present, the study of the large time
behavior is not immediate. In fact, despite conservation of the
total number of agents, we don't have any other additional
conservation equation or entropy dissipation. Although we prove
that the moments do not grow more than exponentially, the
determination of an explicit form of the asymptotic wealth
distribution of the kinetic equation remains difficult and
requires the use of suitable numerical methods.

A complementary method to extract information on the tails is
linked to the possibility to obtain particular asymptotics which
maintain the characteristics  of the solution to the original
problem for large times. Following the analysis in \cite{CPT}, we
shall prove that the Boltzmann model converges in a suitable
asymptotic limit towards a convection-diffusion equation of
Fokker-Planck type for the distribution of wealth among
individuals. Other Fokker-Planck equations were obtained using
different approaches in \cite{BM, So, SMBSR}.

In this case, however, we can show that the Fokker-Planck equation
admits self-similar solutions that can be computed explicitly and
which are lognormal distributions. One is then led to the
conclusion that the formation of Pareto tails in the wealth
distribution observed in \cite{BM, CPT} is a consequence of the
interplay between the conservative binary exchanges having the
effect of redistributing wealth among agents and the speculative
trading causing the growth of mean wealth and social
inequalities.

%In a
%similar way we can also derive a Fokker-Planck equation for the
%distribution of the log-returns of the price.

The rest of the paper is organized as follows. In the next
section, we introduce briefly the microscopic dynamic of the LLS
model. The mesoscopic model is then derived in Section 3 and its
properties discussed in Section 4. These properties justify the
asymptotic procedures performed in section 5. The model behavior
together with its asymptotic limit is illustrated by several
numerical results in section 6. Some conclusions and remarks on
future developments are then made in the last section.

\section{The microscopic dynamic}

Let us consider a set of financial agents $i=1,\ldots,N$ who can
create their own portfolio between two alternative investments: a
stock and a bond. We denote by $w_i$ the wealth of agent $i$ and
by $n_i$ the number of stocks of the agent. Additionally we use
the notations $S$ for the price of the stock and $n$ for the total
number of stocks.

The essence of the dynamic is the choice of the agent's portfolio.
More precisely, at each time step each agent selects which
fraction of wealth to invest in bonds and which fraction in
stocks. We indicate with $r$ the (constant) interest rate of
bonds. The {bond} is assumed to be a risk-less asset yielding a
return at the end of each time period. The {stock} is a risky
asset with overall returns rate $x$ composed of two elements: a
capital gain or loss and the distribution of dividends.

To simplify the notation, let us neglect for the moment the
effects due to the stochastic nature of the process, the presence
of dividends, and so on. Thus, if an agent has invested $\gamma_i
w_i$ of its wealth in stocks and $(1-\gamma_i)w_i$ of its wealth
in bonds, at the next time step in the dynamic he will achieve the
new wealth value
\begin{equation}
w_i'=(1-\gamma_i)w_i(1+r)+\gamma_iw_i (1+x), \label{eq0}
\end{equation}
where the rate of return of the stock is given by
\begin{equation}
x=\frac{S'-S}{S},
\end{equation}
and $S'$ is the new price of the stock.

Since we have the identity
\begin{equation}
\gamma_i w_i = n_i S, \label{eq1}
\end{equation}
we can also write
\begin{eqnarray}
w_i'&=&w_i+w_i(1-\gamma_i)r+w_i\gamma_i\left(\frac{S'-S}{S}\right)\\
&=&w_i+(w_i-n_i S)r+n_i(S'-S).
\end{eqnarray}
Note that, independently of the number of stocks of the agent at
the next time level, it is only the price variation of the stock
(which is unknown) that characterizes the gain or loss of the
agent on the stock market at this stage.

The dynamic now is based on the agent choice of the new fraction
of wealth he wants to invest in stocks at the next stage. Each
investor $i$ is confronted with a decision where the outcome is
uncertain: which is the new optimal fraction $\gamma_i'$ of wealth
to invest in stock? According to the standard theory of investment
each investor is characterized by a {\it utility function} (of its
wealth) $U(w)$ that reflects the personal risk taking preference
\cite{In}. The optimal $\gamma_i'$ is the one that maximizes the
expected value of $U(w)$.

Different models can be used for this (see \cite{LLS1, VT}), for
example, maximizing a von Neumann-Morgenstern utility function
with a constant risk aversion of the type
\begin{equation}
U(w)=\frac{w^{1-\alpha}}{1-\alpha}, \label{equf}
\end{equation}
where $\alpha$ is the risk aversion parameter, or a logarithmic
utility function \begin{equation} U(w)=\log(w).\end{equation}

As they don't know the future stock price $S'$, the investors
estimate the stock's next period return distribution and find an
optimal mix of the stock and the bond that maximizes their
expected utility $E[U]$. In practice, for any hypothetical price
$S^h$, each investor finds the hypothetical optimal proportion
$\gamma_i^h(S^h)$ which maximizes his/her expected utility
evaluated at
\begin{equation}
w_i^h(S^h)=(1-\gamma_i^h)w_i'(1+r)+\gamma_i^hw_i' (1+x'(S^h)),
\end{equation}
where $x'(S^h)=(S^h-S')/S'$ and $S'$ is estimated in some way. For
example in \cite{LLS1} the investors expectations for $x'$ are
based on extrapolating the past values.

Note that, if we assume that all investors have the same risk
aversion $\alpha$ in (\ref{equf}), then they will have the same
proportion of investment in stocks regardless of their wealth,
thus $\gamma_i^h(S^h)=\gamma^h(S^h)$.

Once each investor decides on the hypothetical optimal proportion
of wealth $\gamma_i^h$ that he/she wishes to invest in stocks, one
can derive the number of stocks $n^h_i(S^h)$ he/she wishes to hold
corresponding to each hypothetical stock price $S^h$. Since the
total number of shares in the market $n$, is fixed there is a
particular value of the price $S'$ for which the sum of the
$n_i^h(S^h)$ equals $n$. This value $S'$ is the new market
equilibrium price and the optimal proportion of wealth is
$\gamma_i'=\gamma_i^h(S')$.

More precisely, following \cite{LLS1}, each agent formulates a
{\it demand curve}
\[n^h_i=n^h_i(S^h)=\frac{\gamma^h(S^h)w_i^h(S^h)}{S^h}\] characterizing the desired
number of stocks as a function of the hypothetical stock price
$S^h$. This number of share demands is a monotonically decreasing
function of the hypothetical price $S^h$. As the total number of
stocks
\begin{equation}
n=\sum_{i=1}^N n_i \label{eq3}
\end{equation}
is preserved, the new price of the stock at the next time level is
given by the so-called {\it market clearance condition}. Thus the
new stock price $S'$ is the unique price at which the total demand
equals the supply
\begin{equation}
\sum_{i=1}^N n_i^h(S')=n. \label{eqmk}
\end{equation}
This will fix the value $w'$ in (\ref{eq0}) and the model can be
advanced to the next time level. To make the model more realistic,
typically a source of stochastic noise, which characterizes all
factors causing the investor to deviate from his/her optimal
portfolio, is introduced in the proportion of investments
$\gamma_i$ and in the rate of return of the stock $x'$.

\section{Kinetic modelling}
We define $f=f(w,t)$, $w \in \mathbb{R_+}$, $t>0$ the distribution
of wealth $w$, which represents the probability for an agent to
have a wealth $w$. We assume that at time $t$ the percentage of
wealth invested is of the form $\gamma(\xi)=\mu(S)+\xi$, where
$\xi$ is a random variable in $[-z,z]$, and
$z=\min\{-\mu(S),1-\mu(S)\}$ is distributed according to some
probability density $\Phi(\mu(S),\xi)$ with zero mean and variance
$\zeta^2$. This probability density characterizes the individual
strategy of an agent around the optimal choice $\mu(S)$. We
assume $\Phi$ to be independent of the wealth of the agent. Here,
the optimal demand curve $\mu(\cdot)$ is assumed to be a given
monotonically non-increasing function of the price $S\geq 0$ such
that $0<\mu(0)<1$.

\begin{figure}
\begin{center}
\includegraphics[scale=.45]{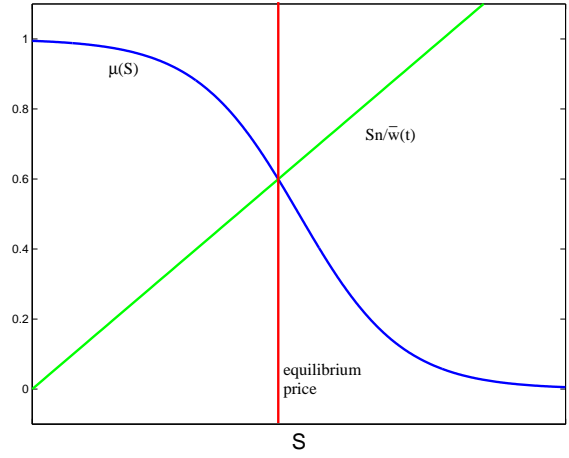}
\end{center}
\caption{Example of equilibrium price} \label{fgequilibrium}
\end{figure}

Note that given $f(w,t)$ the actual stock price $S$ satisfies the
demand-supply relation
\begin{equation}
S=\frac1{n} E[\gamma w], \label{eqexp}
\end{equation}
where $E[X]$ denotes the mathematical expectation of the random
variable $X$ and $f(w,t)$ has been normalized
\[
\int_0^\infty f(w,t) dw=1.
\]
More precisely, since $\gamma$ and $w$ are independent, at each
time $t$, the price $S(t)$ satisfies (see Figure
\ref{fgequilibrium})
\begin{equation}
S(t)=\frac1{n} E[\gamma]E[w]=\frac1{n}\mu(S(t)){\bar w}(t),
\label{eqS}
\end{equation} with
\begin{equation}
{\bar w}(t)\stackrel{def}{:=}E[w]=\int_0^\infty f(w,t) w dw
\end{equation}
being the mean wealth and by construction,
\[
\mu(S)=\int \Phi(\mu(S),\xi)\xi\,d\xi.
\]

At the next round in the market, the new wealth of the investor
will depend on the future price $S'$ and the percentage $\gamma$ of wealth
invested according to
\begin{equation}
w'(S',\gamma,\eta)=(1-\gamma)w(1+r)+\gamma w (1+x(S',\eta)),
\label{eqw1}
\end{equation}
where the expected rate of return of stocks is given by
\begin{equation}
x(S',\eta)=\frac{S'-S+D+\eta}{S}. \label{eqx1}
\end{equation}
In the above relation, $D\geq 0$ represents a constant dividend
paid by the company and $\eta$ is a random variable distributed
according to $\Theta(\eta)$ with zero mean and variance
$\sigma^2$, which takes into account fluctuations due to price
uncertainty and dividends \cite{LLS1, HT01}. We assume $\eta$ to
take values in $[-d,d]$ with $0< d\leq S'+D$ so that $w' \geq 0$
and thus negative wealths are not allowed in the model. Note that
equation (\ref{eqx1}) requires estimation of the future price $S'$,
which is unknown.

The dynamic is then determined by the agent's new fraction of
wealth invested in stocks, $\gamma'(\xi')=\mu(S')+\xi'$, where
$\xi'$ is a random variable in $[-z',z']$ and
$z'=\min\{\mu(S'),1-\mu(S')\}$ is distributed according to
$\Phi(\mu(S'),\xi')$. We have the demand-supply relation
\begin{equation}
S'=\frac1{n} E[\gamma' w'],
\end{equation}
which permits us to write the following equation for the future price:
\begin{eqnarray}
S'=\frac1{n} E[\gamma']E[w']=\frac1{n} \mu(S')E[w'].
\label{eq:newpr}
\end{eqnarray}
Now
\begin{equation}
w'(S',\gamma,\eta)=w(1+r)+\gamma w  (x(S',\eta)-r), \label{eq00}
\end{equation}
thus
\begin{eqnarray}
E[w']&=&E[w](1+r)+E[\gamma w](E[x(S',\eta)]-r)\\
&=&{\bar w}(t)(1+r)+\mu(S){\bar
w}(t)\left(\frac{S'-S+D}{S}-r\right).
\end{eqnarray}
This gives the identity
\begin{eqnarray}
S'=\frac1{n} \mu(S'){\bar
w}(t)\left[(1+r)+\mu(S)\left(\frac{S'-S+D}{S}-r\right)\right].
\end{eqnarray}
Using equation (\ref{eqS}) we can eliminate the dependence on the
mean wealth and write
\begin{eqnarray}
\nonumber S'&=&\frac{\mu(S')}{\mu(S)}
\left[(1-\mu(S))S(1+r)+\mu(S)(S'+D)\right]\\
&=&
\frac{(1-\mu(S))\mu(S')}{(1-\mu(S'))\mu(S)}(1+r)S+\frac{\mu(S')}{1-\mu(S')}D.
\label{eqprice}
\end{eqnarray}

\begin{remark}{˜}\\\rm
The equation for the future price deserves some remarks.
\begin{itemize}
\item Equation (\ref{eqprice}) determines implicitly the future
value of the stock price. Let us set
\beq{g(S)=\frac{1-\mu(S)}{\mu(S)}S.} Then the future price is
given by the equation \beq{g(S')=g(S)(1+r)+D} for a given $S$.
Note that
\[
\frac{d g(S)}{d S}=-\frac{d \mu(S)}{d
S}\frac{S}{\mu(S)^2}+\frac{1-\mu(S)}{\mu(S)}>0,
\]
so the function $g(S)$ is strictly increasing with respect
to $S$. This guarantees the existence of a unique solution
\begin{equation}
S'=g^{-1}\left(g(S)(1+r)+D\right)>S.
\end{equation}
Moreover, if $r=0$ and $D=0$, the unique solution is $S'=S$ and the
price remains unchanged in time.

For the average stock return, we have \be {\bar x}(S') - r =
\frac{(\mu(S')-\mu(S))(1+r)}{(1-\mu(S'))\mu(S)}+
\frac{\mu(S')D}{S(1-\mu(S'))}, \label{eqxr} \ee where \be {\bar
x}(S')=E[x(S',\eta)]=\frac{S'-S+D}{S}. \ee Now the right hand side
of (\ref{eqxr}) has non-constant sign since $\mu(S')\leq \mu(S)$.
In particular, the average stock return is above the bonds rate
$r$ only if the (negative) rate of variation of the investments is
above a certain threshold
\[
\frac{\mu(S')-\mu(S)}{\mu(S)\mu(S')}S \geq -\frac{D}{(1+r)}.
\]
\item In the constant investment case $\mu(\cdot)=C$, with $C\in (0,1)$ constant,
then we have $g(S)=(1-C)S/C$ and \beq{S'=(1+r)S+\frac{C}{1-C}D,}
which corresponds to a dynamic of growth of the prices at rate $r$.
As a consequence, the average stock return is always larger then
the constant return of bonds:
\[
{\bar x}(S') - r = \frac{D}{S(1-C)} \geq 0.
\]
\end{itemize}
\end{remark}

%Let us first restrict to the case $\mu(\cdot)=\mu(\cdot)$.
By standard methods of kinetic theory \cite{CIP}, the microscopic
dynamics of agents originate the following linear kinetic equation
for the evolution of the wealth distribution
\begin{equation}
\frac{\partial f(w,t)}{\partial t}=\int_{-d}^{d} \int_{-z}^{z}
\left(\beta('w\to w)\frac1{j(\xi,\eta,t)}f('w,t)- \beta(w\to
w')f(w,t)\right)d\xi\,d\eta. \label{eqkin}
\end{equation}
The above equation takes into account all possible variations
that can occur to the distribution of a given wealth $w$. The
first part of the integral on the right hand side takes into
account all possible gains of the test wealth $w$ coming from a
pre-trading wealth $'w$. The function $\beta('w\to w)$ gives the
probability per unit time of this process.

Thus $'w$ is obtained simply by inverting the dynamics to get
\begin{equation}
'w=\frac{w}{j(\xi,\eta,t)},\quad
j(\xi,\eta,t)=1+r+\gamma(\xi)(x(S',\eta)-r),
\end{equation}
where the value $S'$ is given as the unique fixed point of
(\ref{eq:newpr}).

The presence of the term $j$ in the integral is needed in order to
preserve the total number of agents
\[
\frac{d}{dt}\int_0^\infty f(w,t)dw=0.
\]
The second part of the integral on the right hand side of
(\ref{eqkin}) is a negative term that takes into account all
possible losses of wealth $w$ as a consequence of the direct
dynamic (\ref{eqw1}), the rate of this process now being
$\beta(w\to w')$. In our case, the kernel $\beta$ takes the form
\begin{eqnarray}
\beta(w\to w')=\Phi(\mu(S),\xi)\Theta(\eta).
\end{eqnarray}
 The distribution function $\Phi(\mu(S),\xi)$, together with
 the function $\mu(\cdot)$, characterizes the behavior of the agents on the market (more
precisely, they characterize the way the agents invest their wealth
as a function of the actual price of the stock).

\begin{remark}\rm
In the derivation of the kinetic equation, we assumed for
simplicity that the actual demand curve $\mu(\cdot)$ which gives
the optimal proportion of investments is a function of the price
only. In reality, the demand curve should change at each market
iteration and should thus depend also on time. In the general case
where each agent has a wealth-dependent individual strategy, one
should consider the distribution $f(\gamma,w,t)$ of agents having
a fraction $\gamma$ of their wealth $w$ invested in stocks.
\end{remark}

\section{Properties of the kinetic equation}

We will start our  analysis  by introducing some notations. Let
$\bigM_0$ be the space of all probability measures on $\real_+$ and
by \bq\label{misure} \bigM_{p} =\left\{ \Psi \in\bigM_0:
\int_{\real_+} |\vartheta|^{p}\Psi(\vartheta)\, d\vartheta <
+\infty, p\ge 0\right\}, \eq we mean the space of all Borel probability
measures of finite momentum of order $p$, equipped with the
topology of the weak convergence of the measures.

Let $\bigF_p(\real_+)$, $p>1$  be the class of all real  functions
on $\real_+$ such that $g(0)= g'(0) =0$ and $g^{(m)}(v)$ is
H\"older continuous of order $\delta$,
 \bq\label{lip} \|g^{(m)}\|_\delta= \sup_{v\not= w} \frac{|g^{(m)}(v) -g^{(m)}(w)|}{
|v-w|^\delta} <\infty,
 \eq
 the integer $m$ and the number $0 <\delta \le 1$ be such that $m+\delta =p$, and
$g^{(m)}$ denote the $m$-th derivative of $g$.

Clearly the symmetric probability
 density $\Theta$ which characterizes the stock returns belongs to
$\bigM_{p}$ for all $p>0$ since
\[
\int_{-d}^{d}|\eta|^{p}\Theta(\eta) d\eta \leq |d|^p.
\]
Moreover, to simplify  computations, we assume that this density
is obtained
 from a given random variable $Y$ with zero mean and unit
 variance. Thus $\Theta$ of variance $\sigma^2$ is the density of
 ${\sigma}Y$. By this assumption, we can easily obtain the dependence on $\sigma$ of the moments of
$\Theta$.
 In fact, for any $p >2$,
 \[
\int_{-d}^{d}|\eta|^{p}\Theta(\eta) d\eta = E\left(
\left|{\sigma}Y\right|^{p}\right) =
\sigma^{p}E\left(\left|Y\right|^{p}\right).
\]

Note that equation (\ref{eqkin}) in weak form takes the simpler
form
\begin{eqnarray}
\frac{d}{dt}\int_0^\infty
f(w,t)\phi(w)dw=\int_0^\infty\int_{-D}^{D}\int_{-z}^{z}
\Phi(\mu(S),\xi)\Theta(\eta)
f(w,t)(\phi(w')-\phi(w))d\xi\,d\eta\,dw. \label{eqkinw}
\end{eqnarray}
By a weak solution of the initial value problem for equation
(\ref{eqkin}) corresponding to the initial probability density
$f_0(w) \in \bigM_{p}$, $p >1$, we shall mean any probability density
$f \in C^1(\real_+, \bigM_{p})$ satisfying the weak form
(\ref{eqkinw}) for $t>0$ and all $\phi \in \bigF_{p}(\real_+)$,
and such that for all $\phi \in \bigF_{p}(\real_+)$,
 \bq\label{ic} \lim_{t\to 0} \int_{0}^{\infty} f(w,t)\phi(w)\, dw = \int_{0}^{\infty} f_0(w)\phi(w)\, dw.
  \eq
The form (\ref{eqkinw}) is easier to handle, and it is the
starting point to study the evolution of macroscopic quantities
(moments). The existence of a weak solution to equation
(\ref{eqkin}) can be seen easily using the same methods available
for the linear Boltzmann equation (see \cite{ST} and the
references therein for example).

From (\ref{eqkinw}) follows the conservation of the total number
of investors if $\phi(w)=1$. The choice  $\phi(w)= w$ is of
particular interest since it gives the time evolution of the
average wealth which characterizes the price behavior. In fact,
the mean wealth is not conserved in the model since we have
\begin{eqnarray}
\frac{d}{dt}\int_0^\infty
f(w,t)w\,dw=\left(r+\mu(S)\left(\frac{S'-S+D}{S}-r\right)\right)\int_0^\infty
f(w,t)w\,dw. \label{eqkinww}
\end{eqnarray}
Note that since the sign of the right hand side is nonnegative, the
mean wealth is nondecreasing in time. In particular, we can rewrite
the equation as
\begin{eqnarray}
\frac{d}{dt}{\bar w(t)}=\left((1-\mu(S))r+\mu(S){\bar
x}(S')\right){\bar w(t)}. \label{eqkinww2}
\end{eqnarray}
From this we get the equation for the price \be
\frac{d}{dt}{S(t)}=\frac{\mu(S(t))}{\mu(S(t))-\dot{\mu}(S(t))S(t)}\left((1-\mu(S(t)))r+\mu(S(t)){\bar
x}(S'(t))\right)S(t), \label{eqkinpr} \ee where $S'$ is given by
(\ref{eqprice}) and
\[
\dot{\mu}(S)=\frac{d\mu(S)}{d S}\leq 0.
\]
Now since from (\ref{eqxr}) it follows by the monotonicity of
$\mu$ that
\[
{\bar x}(S') \leq M {\stackrel{def}{:=}}
r+\frac{D}{S(0)(1-\mu(S(0)))},
\]
using (\ref{eqkinww2}) we have the bound
\begin{eqnarray}
{\bar w(t)}\leq {\bar w}(0) \exp\left(Mt\right). \label{eqwbound}
\end{eqnarray}
From (\ref{eqS}) we obtain immediately
\[
\frac{S(t)}{\mu(S(t))}\leq \frac{S(0)}{\mu(S(0))}
\exp\left(Mt\right),
\]
which gives
\begin{eqnarray}
{S(t)}\leq {S(0)} \exp\left(Mt\right). \label{eqSbound}
\end{eqnarray}
\begin{remark}\rm
For a constant $\mu(\cdot)=C$, $C\in(0,1)$ we have the explicit
expression for the growth of the wealth (and consequently of the
price) \be {\bar w}(t)={\bar
w}(0)\exp(rt)-(1-\exp(rt))\frac{nD}{1-C}. \label{eq:muc}\ee
\end{remark}

Analogous bounds to (\ref{eqwbound}) for  moments of  higher order
can be obtained in a similar way. Let us consider the case of
moments of order $p\geq 2$, which we will need in the sequel.
Taking $\phi(w) = w^p$, we get
 \be
\frac d{dt}\int_{0}^\infty w^p f(w,t)\,dw  =
\int_0^\infty\int_{-d}^{d}\int_{-z}^{z}
\Phi(\mu(S),\xi)\Theta(\eta) f(w,t)(w'^p-w^p)d\xi\,d\eta\,dw. \ee
Moreover, we can write
 \[
 w'^p = w^p + pw^{p-1}(w' - w) + \frac 12 p(p-1)\tilde w^{p-2}(w'-w)^2,
 \]
 where, for some $0 \le \vartheta \le 1$,
 \[
 \tilde w = \vartheta w' +(1 -\vartheta)w .
  \]
Hence,
\[
\int_0^\infty\int_{-d}^{d}\int_{-z}^{z}
\Phi(\mu(S),\xi)\Theta(\eta) f(w,t)(w'^p-w^p)d\xi\,d\eta\,dw
\]
\[=\int_0^\infty\int_{-d}^{d}\int_{-z}^{z}
\Phi(\mu(S),\xi)\Theta(\eta) f(w,t)(pw^{p-1}(w' - w) +\frac 12
p(p-1)\tilde w^{p-2}(w'-w)^2 )d\xi\,d\eta\,dw \]
\[ = p((1-\mu(S))r+\mu(S){\bar
x}(S'))\int_{0}^\infty w^p f(w,t)\,dw+ \frac 12 p(p-1)\]
\[
\int_0^\infty\int_{-d}^{d}\int_{-z}^{z}
\Phi(\mu(S),\xi)\Theta(\eta) f(w,t)\tilde w^{p-2}w^2((1-\gamma)
r+\gamma{x}(S',\eta))^2d\xi\,d\eta\,dw.
\]
From
\[
\tilde w^{p-2}=w^{p-2}(1+\vartheta((1-\gamma)r+\gamma{
x}(S',\eta)))^{p-2}\leq w^{p-2}(1+r+ |{ x}(S',\eta)|)^{p-2}
\]
\[
\leq C_pw^{p-2}(1+r^{p-2}+ |{x}(S',\eta)|^{p-2})
\]
and
\[
((1-\gamma)r+\gamma{ x}(S',\eta)))^{2}\leq 2(r^2+{
x}(S',\eta)^{2}),
\]
we have
\[
\int_0^\infty\int_{-d}^{d}\int_{-z}^{z}
\Phi(\mu(S),\xi)\Theta(\eta) f(w,t)\tilde w^{p-2}w^2 ((1-\gamma)
r+\gamma{x}(S',\eta))^2 d\xi\,d\eta\,dw
\]
\[
\leq 2C_p \int_0^\infty\int_{-d}^{d} \Theta(\eta)
f(w,t)w^{p}(1+r^{p-2}+|{x}(S',\eta)|^{p-2})(r^2+{
x}(S',\eta)^{2})d\eta\,dw.
\]
Since \be \int_{-d}^{d} \Theta(\eta)|{x}(S',\eta)|^{p}\,d\eta \leq
\frac{c_p}{S^p}\left((S'-S)^p+D^p+\sigma^p
E(|Y|^p)\right)\label{eqeta}, \ee we finally obtain the bound
 \be
\frac d{dt}\int_{0}^\infty w^p f(w,t)\,dw  \leq  A_p(S)
\int_{0}^\infty w^p f(w,t)\,dw, \label{eqwpbound}\ee where
\begin{eqnarray*}
A_p(S)&=& p((1-\mu(S))r+\mu(S){\bar x}(S'))\\
&+& p(p-1)
C_p\left[r^p+(1+r^{p-2})\left(1+\frac{c_2}{S^2}((S'-S)^2+D^2+\sigma^2
E(|Y|^2))\right)\right.\\
&+&r^{2}\left(1+\frac{c_{p-2}}{S^{p-2}}((S'-S)^{p-2}+D^{p-2}+\sigma^{p-2}
E(|Y|^{p-2}))\right)\\
&+&\left.\left(\frac{c_{p}}{S^{p}}((S'-S)^{p}+D^{p}+\sigma^{p}
E(|Y|^{p}))\right)\right]\end{eqnarray*} and $C_p$, $c_p$,
$c_{p-2}$ and $c_2$ are suitable constants.

We can summarize our results in the following

\begin{theorem}
 Let the probability density $f_0 \in \bigM_p$, where $p= 2+ \delta$ for some $\delta>0$. Then the average wealth is
 increasing exponentially with time following (\ref{eqwbound}). As
 a consequence, if $\mu$ is a non-increasing function of $S$, the price does not grow more than
 exponentially as in (\ref{eqSbound}).
Similarly, higher order moments do not increase more than
exponentially, and we have the bound (\ref{eqwpbound}).
\end{theorem}

\section{Fokker-Planck asymptotics and self-similar solution}

The previous analysis shows that in general it is difficult to
study in detail the asymptotic behavior of the system. In
addition, we must take into account the exponential growth of the
average wealth. In this case, one way to get information on the
properties of the solution for large time relies on a suitable
scaling of the solution. As is usual in kinetic theory, however,
particular asymptotics of the equation result in simplified models
(generally of Fokker-Planck type) whose behavior is easier to
analyze. Here, following the analysis in \cite{CPT,
PT1} and inspired by similar asymptotic limits for inelastic
gases \cite{EB2, SL}, we consider the limit of large times in which
the market originates a very small exchange of wealth (small rates
of return $r$ and $x$).

In order to study the asymptotic behavior of the distribution
function $f(w,t)$, we start from the weak form of the kinetic
equation
\begin{eqnarray}
\frac{d}{dt}\int_0^\infty
f(w,t)\phi(w)dw=\int_0^\infty\int_{-d}^{d}\int_{-z}^{z}
\Phi(\mu(S),\xi)\Theta(\eta)
f(w,t)(\phi(w')-\phi(w))d\xi\,d\eta\,dw
\end{eqnarray}
and consider a second-order Taylor expansion of $\phi$ around $w$,
 $$
\phi(w') - \phi(w) = w(r+\gamma (x(S',\eta)- r)) \phi'(w) + {1
\over 2} w^2(r+\gamma (x(S',\eta)- r))^2 \phi''(\tilde w),
 $$
where, for some $0 \le \vartheta \le 1$,
 \[
 \tilde w = \vartheta w' +(1 -\vartheta)w .
  \]
 Inserting this expansion into the collision operator,  we get
\ba \nonumber &&\int_0^\infty\int_{-d}^{d}\int_{-z}^{z}
\Phi(\mu(S),\xi)\Theta(\eta)
f(w,t)(\phi(w')-\phi(w))d\xi\,d\eta\,dw\\
\nonumber
 &&=
\int_0^\infty\int_{-d}^{d}\int_{-z}^{z}
\Phi(\mu(S),\xi)\Theta(\eta) f(w,t)
w(r+\gamma (x(S',\eta)- r)) \phi'(w) d\xi\,d\eta\,dw\\
\nonumber &&+\frac1{2} \int_0^\infty\int_{-d}^{d}\int_{-z}^{z}
\Phi(\mu(S),\xi) \Theta(\eta)f(w,t) w^2(r+\gamma (x(S',\eta)-
r))^2 \phi''(w) d\xi\,d\eta\,dw\\
\nonumber &&+R_r(S,S'), \ea where \ba \nonumber
R_r(S,S')&=&\frac12\int_0^\infty\int_{-d}^{d}\int_{-z}^{z}
\Phi(\mu(S),\xi)\Theta(\eta) f(w,t)\\[-.25cm]
\\
\nonumber &&w^2(r+\gamma (x(S',\eta)- r))^2 (\phi''(\tilde
w)-\phi''(w))\,d\xi\,d\eta\,dw. \ea Recalling that $E[\xi]=0$,
$E[\eta]=0$, $E[\xi^2]=\zeta^2$ and $E[\eta^2]=\sigma^2$, we can
simplify the above expression to obtain \ba \nonumber
&&\int_0^\infty\int_{-d}^{d}\int_{-z}^{z}
\Phi(\mu(S),\xi)\Theta(\eta)
f(w,t)(\phi(w')-\phi(w))d\xi\,d\eta\,dw\\
\nonumber
 &&=
\int_0^\infty f(w,t)
w\left(r+\mu(S)\left(\frac{S'-S+D}{S}- r\right)\right) \phi'(w) \,dw\\
\nonumber &&+\frac1{2} \int_0^\infty f(w,t)
w^2\left(r^2+(\zeta^2+\mu(S)^2)\left(\frac{(S'-S)^2}{S^2}+
\frac{\sigma^2+D^2}{S^2}+2D\frac{S'-S}{S^2}\right.\right.\\
\nonumber &&+\left.\left.r^2-2r\frac{S'-S+D}{S}\right)+
2r\mu(S)\left(\frac{S'-S+D}{S}-r\right)\right)\phi''(w)\,dw\\
\nonumber&&+R_r(S,S'). \ea
Now we set
\[
\tau=rt,\quad \ft(w,\tau)=f(w,t),\quad \St(\tau)=S(t),\quad
\mut(\St)=\mu(S),
\]
which implies that $\ft(w,\tau)$ satisfies the equation
\begin{eqnarray}
\nonumber &&\frac{d}{d\tau}\int_0^\infty \ft(w,\tau)\phi(w)dw\\
\nonumber&&= \int_0^\infty \ft(w,\tau)
w\left(1+\mut(\St)\left(\frac{\St'+D-\St}{r \St}- 1\right)\right) \phi'(w) \,dw\\
\nonumber &&+\frac1{2} \int_0^\infty \ft(w,\tau)
w^2\left(r+(\zeta^2+\mut(\St)^2)\left(\frac{(\St'-\St)^2}{r
\St^2}+
\frac{\sigma^2+D^2}{r\St^2}+2D\frac{\St'-\St}{r\St^2}\right.\right.\\
\nonumber &&+\left.\left.r-2\frac{\St'+D-\St}{\St}\right)+
2\mut(\St)\left(\frac{\St'+D-\St}{\St}-r\right)\right)\phi''(w)\,dw\\
\nonumber &&+\frac1{r}R_r(\St,\St').
\end{eqnarray}
Now we consider the limit of very small values of the constant
rate $r$. In order for such a limit to make sense and preserve the
characteristics of the model, we must assume that \be \lim_{r\to
0}\frac{\sigma^2}{r}=\nu, \quad \lim_{r\to 0}\frac{D}{r}=\lambda.
\label{limits}\ee First let us note that the above limits in
(\ref{eqprice}) imply immediately that \be \lim_{r\to 0} \St'=\St.
\label{limits2}\ee
We begin by showing that the remainder is small for small values of $r$.\\
Since $\phi \in \bigF_{2+\delta}(\real_+)$ and $|\tilde w - w |=
\vartheta|w'-w|$,
 \be\label{rem}
 \left| \phi''(\tilde w)- \phi''( w)\right| \le \| \phi''\|_\delta |\tilde w - w |^\delta \le
 \| \phi''\|_\delta |w' - w |^\delta .
  \ee
  Hence
 \begin{eqnarray*}
|\frac1{r} R_r(\St,\St')| &\le& \frac{\|
\phi''\|_\delta}{2r}\int_0^\infty\int_{-d}^{d}\int_{-z}^{z}
\Phi(\mu(S(\tau)),\xi)\Theta(\eta)
\\&& |(1-\gamma)r+\gamma x(\St',\eta)|^{2+\delta}\ft(w,\tau)w^{2+\delta}  d\xi\,d
\eta\,dw.
 \end{eqnarray*}
By the inequality
 \[
 |(1-\gamma)r+\gamma x(\St',\eta)|^{2+\delta}\leq 2^{2+\delta}
 \left( r^{2+\delta} +|x(\St',\eta)|^{2+\delta}\right)
 \]
and (\ref{eqeta}), we get
 \[
|\frac1{r} R_r(\St,\St')| \leq {2^{1+\delta}\|
\phi''\|_\delta}\cdot \]
\[\cdot\left(r^{1+\delta}+\frac{c_{2+\delta}}{r\St^{2+\delta}}((\St'-\St)^{2+\delta}+D^{2+\delta}
+\sigma^{2+\delta}E(|Y|^{2+\delta}))\right) \int_0^\infty
\ft(w,\tau)w^{2+\delta}  dw.
 \]
As a consequence of (\ref{limits})--(\ref{limits2}), from this
inequality it follows that $R_r(\St,\St')$ converges to zero as $r
\to 0$ if
\[
\int_{0}^{\infty}w^{2+\delta}\ft(w,\tau)\,dw \] remains bounded at
any fixed time $\tau >0$, provided the same bound holds at time
$\tau =0$. This is guaranteed by inequality (\ref{eqwpbound})
since $A_p(\St) \to 0$ in the asymptotic limit defined by
(\ref{limits}).\\
Next we write
\[
\mut(\St')=\mut(\St)+(\St'-\St)\dot{\mut}(\St)+O((\St'-\St)^2),
\]
where
\[
\dot{\mut}(\St)=\frac{d\mut(\St)}{d\St}\leq 0.
\]
Then, using the above expansion from (\ref{limits}) in (\ref{eqprice}),
we obtain \be \lim_{r\to
0}\frac{\St'-\St}{r}=\kappa(\St)\left(\St+\frac{\mut(\St)}{1-\mut(\St)}\lambda\right),
\ee with \be 0 <
\kappa(\St){\stackrel{def}{:=}}\frac{\mut(\St)(1-\mut(\St))}{\mut(\St)(1-\mut(\St))-\St\dot{\mut}(\St)}\leq
1.\label{kappa} \ee Now, sending $r\to 0$ under the same
assumptions, we obtain the weak form
\[
\frac{d}{d\tau}\int_0^\infty \ft(w,\tau)\phi(w)dw\]
\[=\left(1+\mut(\St)\left((\kappa(\St)-1)+
\frac{\mut(\St)(\kappa(\St)-1)+1}{1-\mut(\St)}\frac{\lambda}{\St}\right)\right)\int_0^\infty
\ft(w,\tau) w \phi'(w) \,dw
\]
\[+\frac1{2} \frac{(\mut(\St)^2+\zeta^2)}{\St^2}\nu \int_0^\infty
\ft(w,\tau) w^2\phi''(w)\,dw.
\]
This corresponds to the weak form of the Fokker-Planck equation
\beq{\frac{\partial}{\partial\tau}\ft+A(\tau)
\frac{\partial}{\partial w} (w\ft)=\frac1{2}B(\tau)
\frac{\partial^2}{\partial w^2} (w^2\ft),} or equivalently
\begin{eqnarray}
\frac{\partial}{\partial\tau}\ft&=&\frac{\partial}{\partial
w}\left[-A(\tau)w\ft+ \frac12B(\tau) \frac{\partial}{\partial w}
w^2 \ft\right], \label{eqkinfp}
\end{eqnarray}
with
\begin{eqnarray}
A(\tau)&=&1+\mut(\St)\left((\kappa(\St)-1)+
\frac{\mut(\St)(\kappa(\St)-1)+1}{1-\mut(\St)}\frac{\lambda}{\St}\right)\\
B(\tau)&=&\frac{(\mut(\St)^2+\zeta^2)}{\St^2}\nu.
\end{eqnarray}
Thus we have proved

\begin{theorem}
 Let the probability density $f_0 \in \bigM_p$, where $p= 2+ \delta$ for some $\delta>0$. Then, as $r \to 0$,
 $\sigma \to 0$, and $D\to 0$ in such a way that $\sigma^2 = \nu r$ and $D=\lambda r$, the weak solution
 to the Boltzmann equation (\ref{eqkinw}) for the scaled density
 $\ft_r(w,\tau)=f(v,t)$ with $\tau = r t$
converges, up to extraction of a subsequence, to a probability
density $\ft(w,\tau) $. This density  is a weak solution of the
Fokker-Planck equation (\ref{eqkinfp}).
\end{theorem}

We remark that even for the Fokker-Planck model, the mean wealth is
increasing with time. A simple computation shows that

\begin{eqnarray}
\dot{\bar w}(\tau)=\frac{d}{d\tau}\int_0^\infty \ft(w,\tau) w\,dw
=A(\tau)\int_0^\infty \ft(w,\tau) w\,dw=A(\tau)\bar w(\tau).
\label{eqkinww3}
\end{eqnarray}
Using (\ref{kappa}), we get the bound
\begin{eqnarray}
(1-\mut(\St))\bar w(\tau) +n\lambda \leq \dot{\bar w}(\tau) \leq
\bar w(\tau)+\frac{n\lambda}{1-\mut(\St)}. \label{eqkinww4}
\end{eqnarray}
Similarly, for the second-order moment we have
\begin{eqnarray}
\dot{\bar e}(\tau)=\frac{d}{d\tau}\int_0^\infty \ft(w,\tau)
w^2\,dw =(2A(\tau)+B(\tau))\int_0^\infty \ft(w,\tau)
w^2\,dw=(2A(\tau)+B(\tau))\bar e(\tau). \label{eqkinww5}
\end{eqnarray}

In order to search for self-similar solutions, we consider the
scaling
\[
\ft(w,\tau)=\frac{1}{w}\gt(\chi,\tau),\quad \chi=\log(w).
\]
Simple computations show that $\gt(\chi,\tau)$ satisfies the linear
convection-diffusion equation
\[
\frac{\partial}{\partial
\tau}\gt(\chi,\tau)=\left(\frac{B(\tau)}{2}-A(\tau)\right)
\frac{\partial}{\partial \chi}\gt(\chi,\tau)+\frac{B(\tau)}{2}
\frac{\partial^2}{\partial \chi^2}\gt(\chi,\tau),
\]
which admits the self-similar solution (see \cite{LR} for example)
\begin{equation}
\gt(\chi,\tau)=\frac{1}{(2b(\tau)\pi)^{1/2}}
\exp\left(-\frac{(\chi+{b(\tau)}/{2}-a(\tau))^2}{2b(\tau)}\right),
\end{equation}
where
\[
a(\tau)=\int_0^\tau A(s)\,ds+C_1,\qquad b(\tau)=\int_0^\tau
B(s)\,ds+C_2.
\]
Reverting to the original variables, we obtain the lognormal
asymptotic behavior of the model,
\begin{equation}
\ft(w,\tau)=\frac{1}{w(2b(\tau)\pi)^{1/2}}
\exp\left(-\frac{(\log(w)+{b(\tau)}/{2}-a(\tau))^2}{2b(\tau)}\right).
\label{eq:logn}
\end{equation}
The constants $C_1=a(0)$ and $C_2=b(0)$ can be determined from the
initial data at $t=0$. If we denote by $\bar w(0)$ and $\bar
e(0)$ the initial values of the first two central moments, we get
\[
C_1=\log({\bar w(0)}),\qquad C_2=\log\left(\frac{\bar e(0)}{(\bar
w(0))^2}\right).
\]
Finally, a direct computation shows that
\begin{equation}
a(\tau)=\int_0^\tau \frac{\dot{\bar w}(s)}{\bar
w(s)}\,ds+C_1=\log(\bar w(\tau))
\end{equation}
and
\begin{equation}
b(\tau)=\int_0^\tau \left(\frac{\dot{\bar e}(s)}{\bar
e(s)}-2\frac{\dot{\bar w}(s)}{\bar
w(s)}\right)\,ds+C_2=\log\left(\frac{\bar e(\tau)}{(\bar
w(\tau))^2}\right).
\end{equation}

%with
%\[
%\ft(w,0)=\frac{1}{w(2\pi)^{1/2}}
%\exp\left(-\frac{(\log(w))^2}{B}\right).
%\]

\begin{remark}{~}\rm
\begin{itemize}
\item If we assume $\zeta$ and $\sigma$ are of the same order of magnitude, in
the Fokker-Planck limit the noise introduced by the agents'
deviations with respect to their optimal behavior does not play
any role and the only source of diffusion is due to the stochastic
nature of the returns.
\item In the case of constant investments $\mut(\cdot)=C$, $C\in (0,1)$ we
have the simplified Fokker-Planck equation
\[
\frac{\partial}{\partial\tau}\ft=\frac{\partial}{\partial
w}\left[-\left(1+ \frac{C}{1-C}\frac{\lambda}{\St}\right)w\ft+
\frac12\frac{(C^2+\zeta^2)}{\St^2}\nu \frac{\partial}{\partial w}
w^2 \ft\right].
\]
It is easy to verify that for such a simple situation, the pair
of ordinary differential equations for the evolution of the first
two central moments, (\ref {eqkinww3}) and (\ref{eqkinww5}), can be
solved explicitly.

\item
Imposing the conservation of the mean wealth with the scaling \be
\ft(w,\tau)=\frac{\bar w(0)}{\bar w(\tau)}\ff(v,\tau),\quad
v=\frac{\bar w(0)}{\bar w(\tau)}w \label{eq:wc}, \ee we have the
diffusion equation
\[
\frac{\partial}{\partial \tau}\ff(v,\tau)=\frac{B(\tau)}{2}
\frac{\partial^2}{\partial v^2}(v^2 \ff(v,\tau)).
\]
This yields the asymptotic lognormal behavior
\begin{equation}
\ff(v,\tau)=\frac{1}{v(2\log(\bar E(\tau)/\bar w(0)^2)\pi)^{1/2}}
\exp\left(-\frac{(\log(v)+\log(\sqrt{\bar E(\tau)}/\bar
w(0)))^2}{2\log(\bar E(\tau)/\bar w(0)^2)}\right),
\end{equation}
with
\[
\int_0^\infty \ff(v,\tau) v \,dv=\bar w(0),\quad \bar
E(\tau)=\int_0^\infty \ff(v,\tau) v^2 \,dv.
\]
\end{itemize}
\end{remark}

\section{Numerical examples}
In this section we report the results of different numerical
simulations of the proposed kinetic equations. In all the
numerical tests, we use $N=1000$ agents and $n=10000$ shares.
Initially, each investor has a total wealth of $1000$ composed of
$10$ shares, at a value of $50$ per share, and $500$ in bonds. The
random variables $\xi$ and $\eta$ are assumed distributed
according to truncated normal distributions so that negative
wealth values are avoided (no borrowing and no short selling). In
Tests 1 and 2 we compare the results obtained with the Monte Carlo
simulation of the kinetic model to a direct solution of
the price equation (\ref{eqkinpr}). In the last test case we
consider the time-averaged Monte Carlo asymptotic behavior of the
kinetic model and compare its numerical self-similar solution with
the explicit one computed in the last section using the
Fokker-Planck model.

\subsection*{Test 1}
\begin{figure}
\begin{center}
\includegraphics[scale=.6]{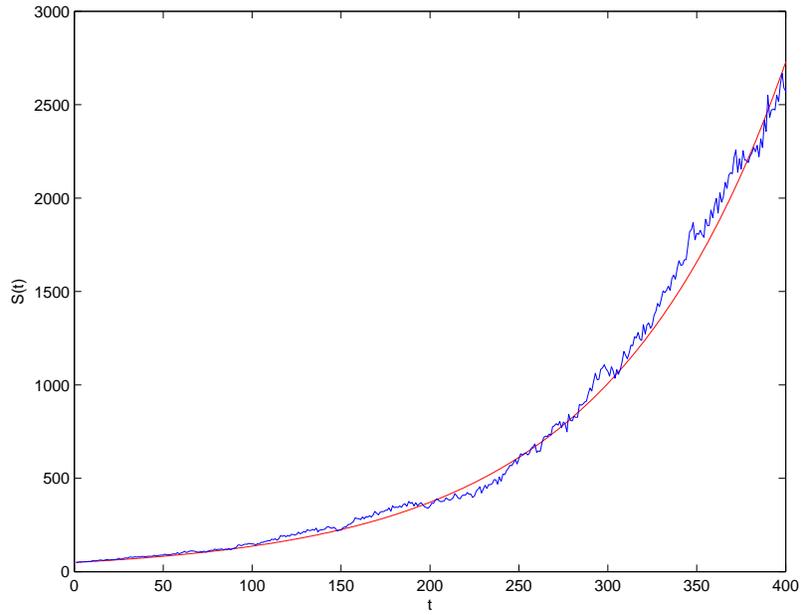}
\end{center}
\caption{Test 1. Exponential growth of the price in time. Numerical
simulation of the kinetic model.} \label{fg:test1}
\end{figure}

\begin{figure}
\begin{center}
\includegraphics[scale=.6]{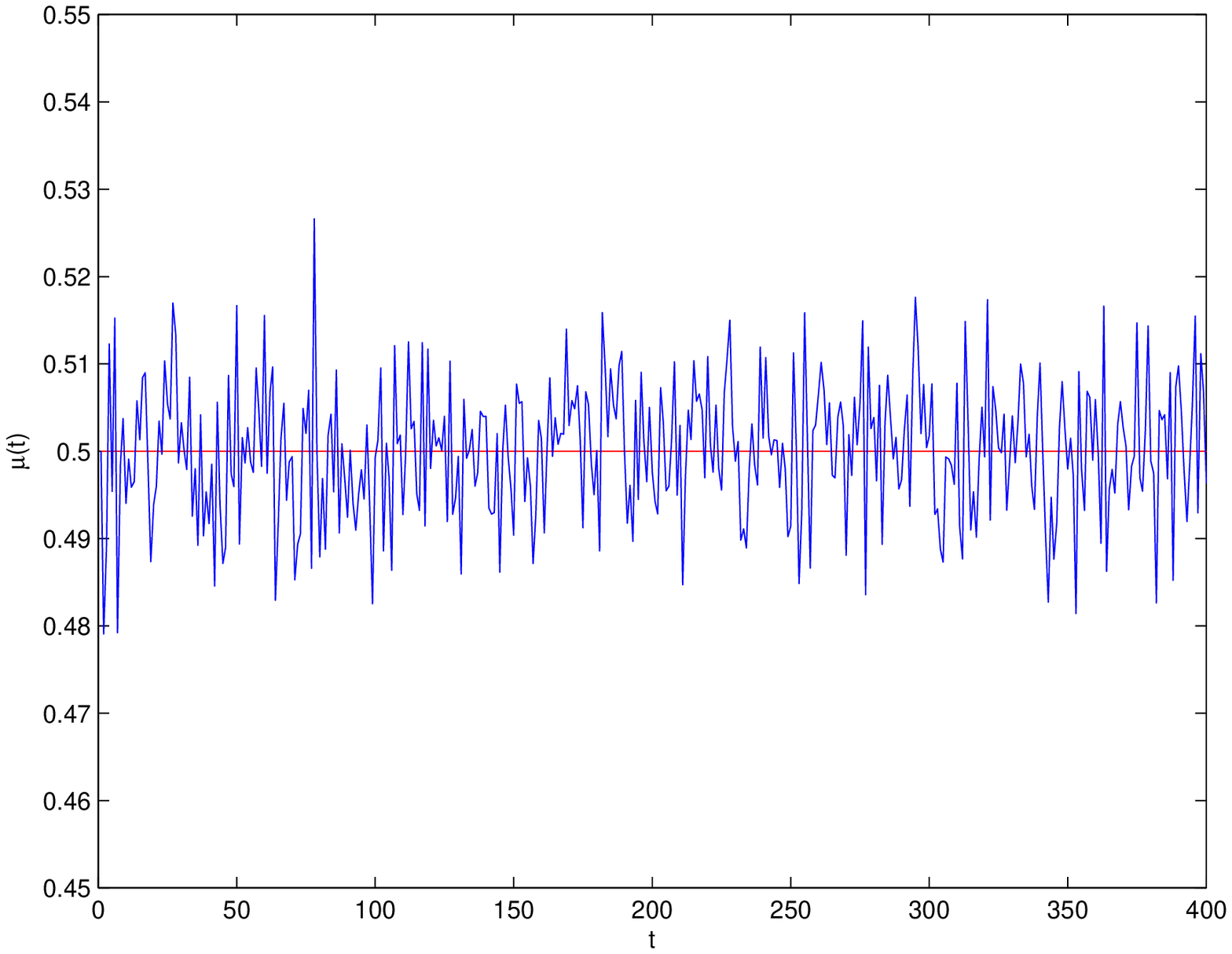}
\end{center}
\caption{Test 1. Fluctuations of the corresponding fraction of
investments in time.} \label{fg:test1b}
\end{figure}

In the first test we take a riskless interest rate $r=0.01$ and an
average dividend growth rate $D=0.015$ and assume that the agents
simply follow a constant investments rule, $\mu(\cdot)=C$, with
$C\in(0,1)$ constant. As a consequence of our choice of parameters
we have $C=0.5$ and the evolution of the mean wealth and of the
price in the kinetic model are known explicitly (\ref{eq:muc}). We
report the results after $400$ stock market iterations with $\xi$
and $\eta/S(0)$ distributed with standard deviation $0.2$ and
$0.3$ respectively. In Figure \ref{fg:test1} we report the
simulated price behavior together with the evolution computed from
(\ref{eqkinpr}). The fraction of investments in time during the Monte
Carlo simulation fluctuates around its optimal value and is given
in Figure \ref{fg:test1b}.

\begin{figure}
\begin{center}
\includegraphics[scale=.6]{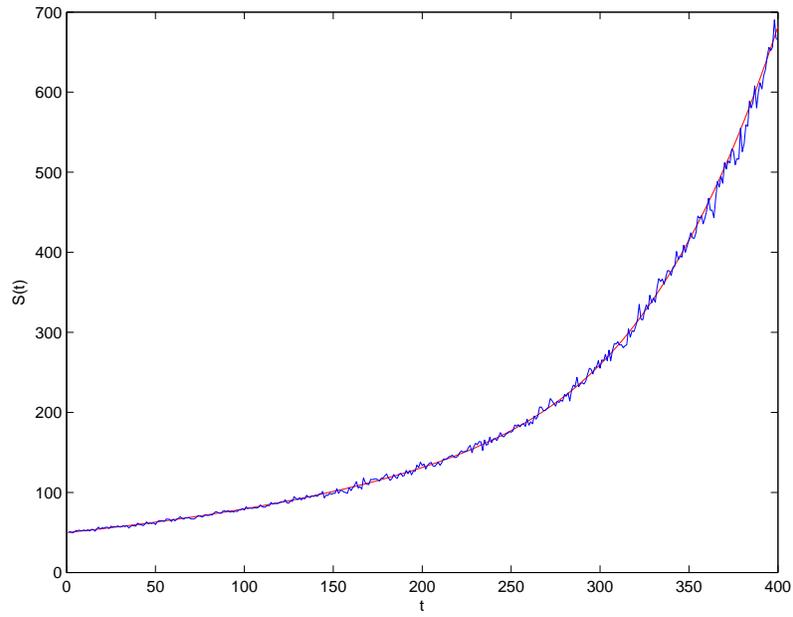}
\end{center}
\caption{Test 2. Price behavior in time. Numerical simulation of
the the kinetic model.} \label{fg:test2}
\end{figure}

\begin{figure}
\begin{center}
\includegraphics[scale=.6]{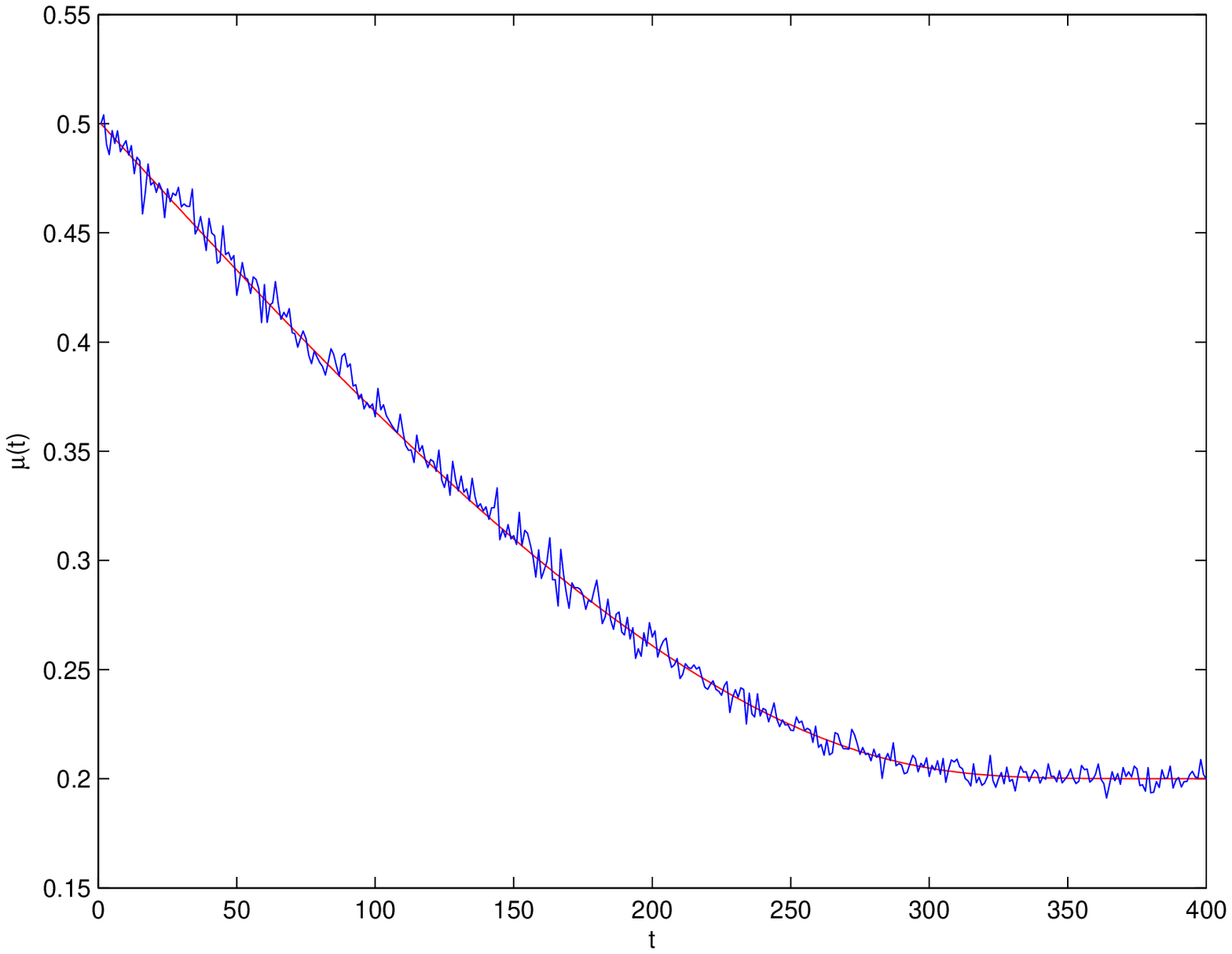}
\end{center}
\caption{Test 2. Fluctuations of the corresponding fraction of
investments in time.} \label{fg:test2b}
\end{figure}

\begin{figure}
\begin{center}
\includegraphics[scale=.6]{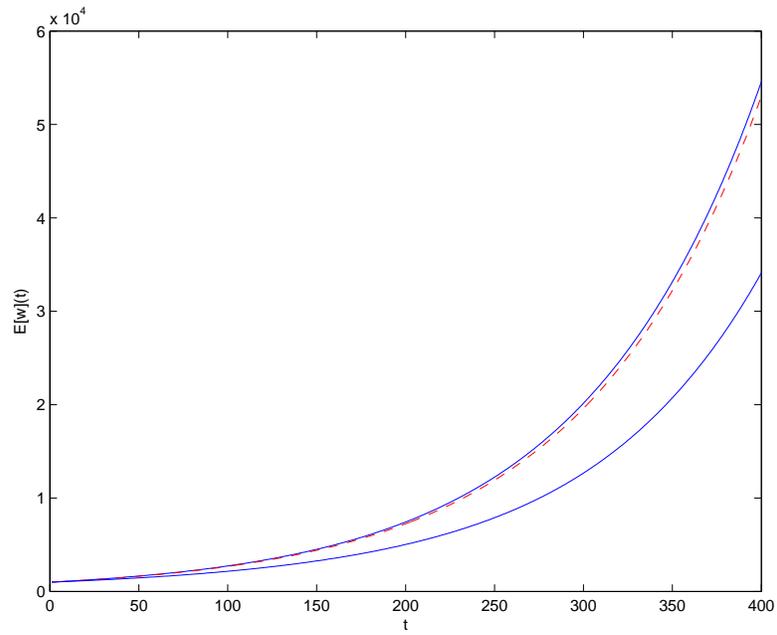}
\end{center}
\caption{Tests 1 and 2. Behavior of the mean wealth in the kinetic
model. The top curve refers to Test 1, the bottom curve to Test 2. The
dashed line corresponds to the exponential growth at a rate equal
to $r$.} \label{fg:test2c}
\end{figure}

\begin{figure}
\begin{center}
\includegraphics[scale=.6]{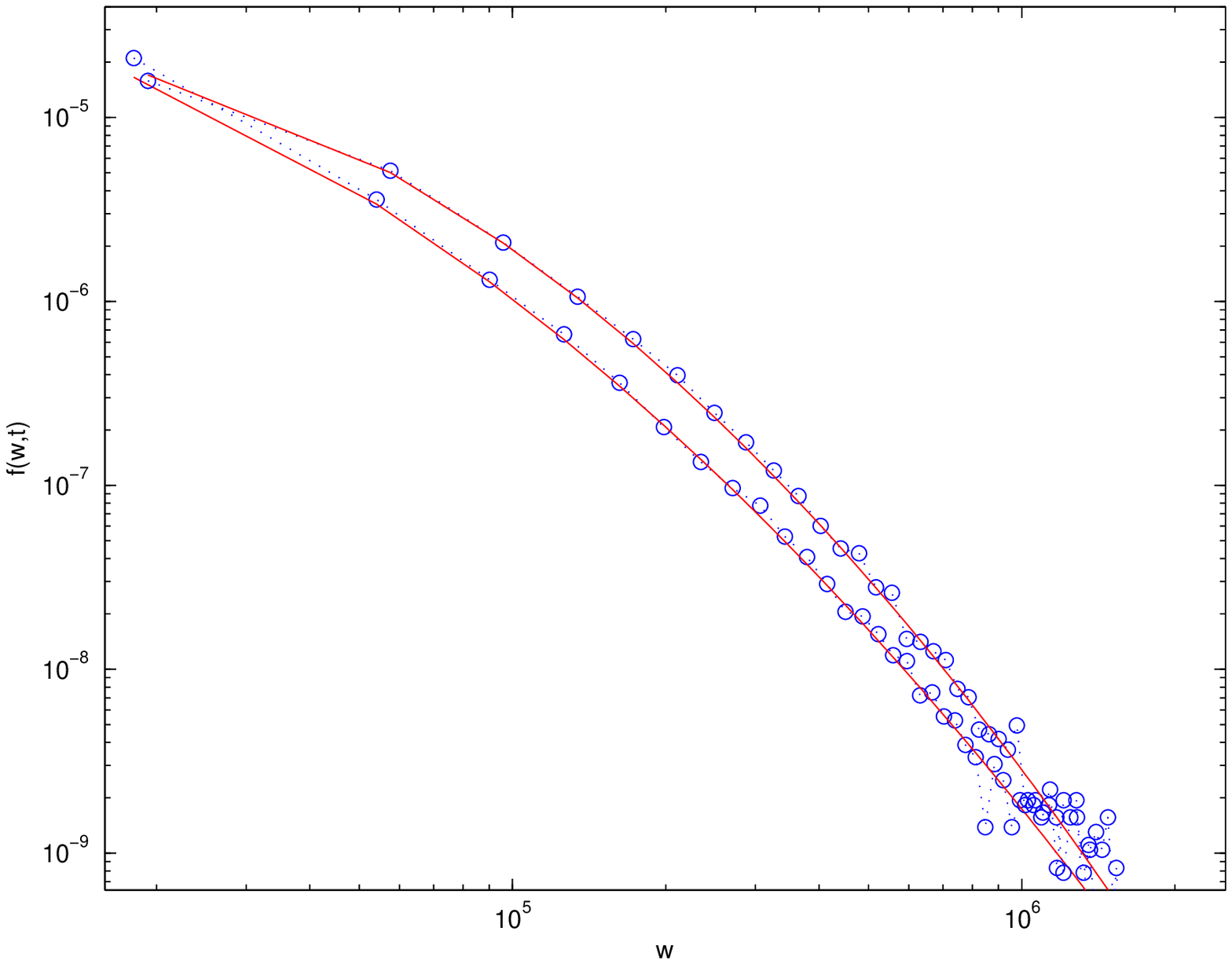}
\end{center}
\caption{Tests 1 and 2. Log-log plot of the mean wealth
distribution together with a lognormal fitting. The top curve refers
to Test 1, the bottom curve to Test 2.} \label{fg:test2d}
\end{figure}

\subsection*{Test 2}
In the next test case we take the same parameters as in Test 1 but
with a non constant profile $\mu(\cdot)$. More precisely we take a
monotone decreasing exponential law
\[
\mu(S)=C_1+(1-C_1)e^{-C_2 S}
\]
with $C_1=0.2$ and $C_2=\log((1-C_1)/(0.5-C_1))/S_0 \approx 0.02$
so that the price equation is satisfied for $S_0=50$. We have
$0.2< \mu(\cdot) \leq 0.5$. The results for the price evolution
and the investments behavior are plotted in Figures \ref{fg:test2}
and \ref{fg:test2b}. The solution for the price in the kinetic
equation has been computed by direct numerical discretization of
(\ref{eqkinpr}). Note that the final price is approximatively $5$
times smaller then the one in the constant investment case with
$\mu=0.5$. In Figure \ref{fg:test2c}, we compare the behavior of
the mean wealth in Tests 1 and 2 to the exponential growth
at a rate $r$ obtained with simple investments in bonds. We can
observe that the time decay of investments in Test 2 is fast
enough to produce a wealth growth below the rate $r$. On the
contrary, as observed in Section 3, a constant investment strategy
produces a curve above this rate. Finally, in Figure
\ref{fg:test2d}, we plot the averaged wealth distributions at the
final computation time on a log-log scale together with a lognormal
fit. The results show  lognormal behavior of the tails even
for the Boltzmann model. Note that thanks to equation (\ref{eq1}),
the same distribution is observed for the number of stocks owned
by the agents.

\subsection*{Test 3}
\begin{figure}
\begin{center}
\includegraphics[scale=.6]{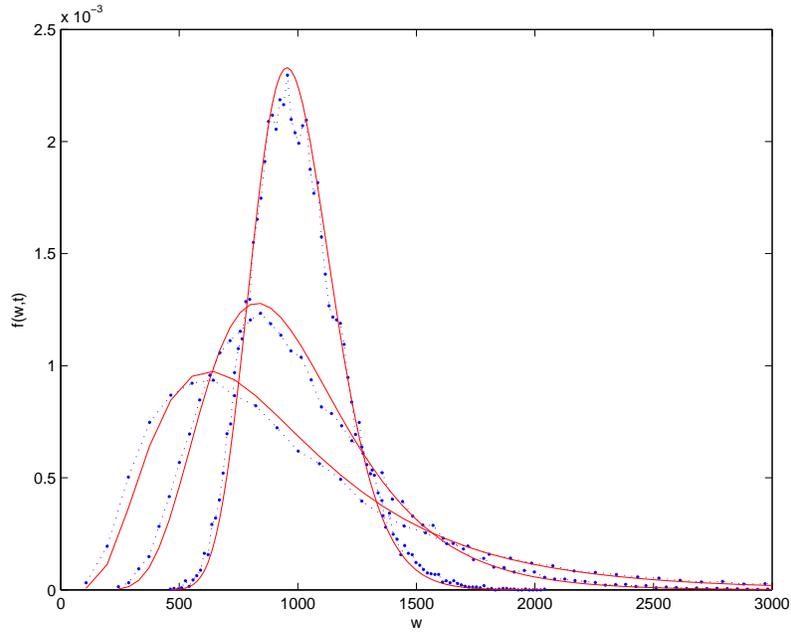}
\end{center}
\caption{Test 3. Distribution function at $t=50,200,500$. The
continuous line is the lognormal Fokker-Planck solution.}
\label{fg:test3}
\end{figure}

\begin{figure}
\begin{center}
\includegraphics[scale=.6]{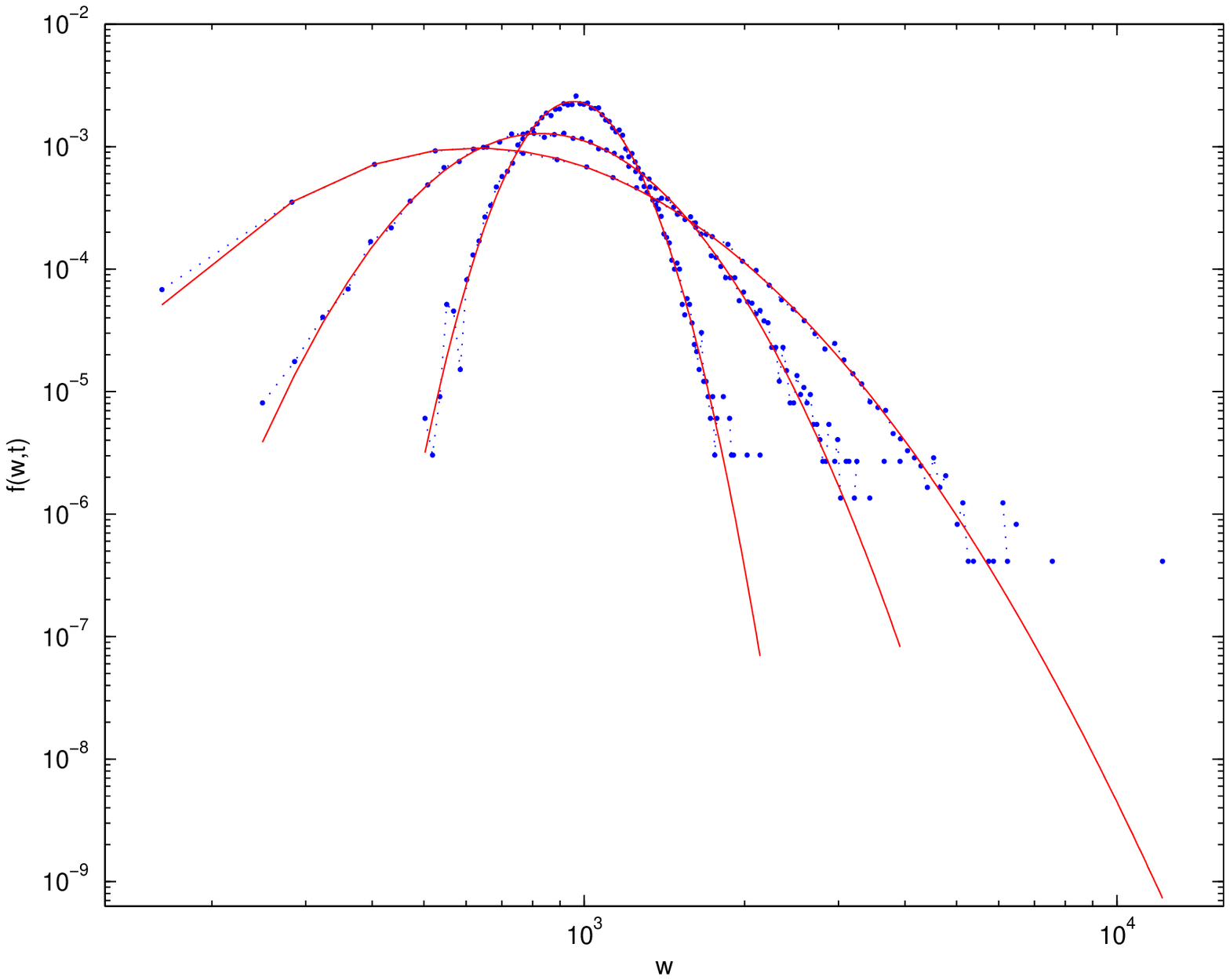}
\end{center}
\caption{Test 3. Log-log plot of the distribution function at
$t=50,200,500$. The continuous line is the lognormal Fokker-Planck
solution.} \label{fg:test3c}
\end{figure}

\begin{figure}
\begin{center}
\includegraphics[scale=.6]{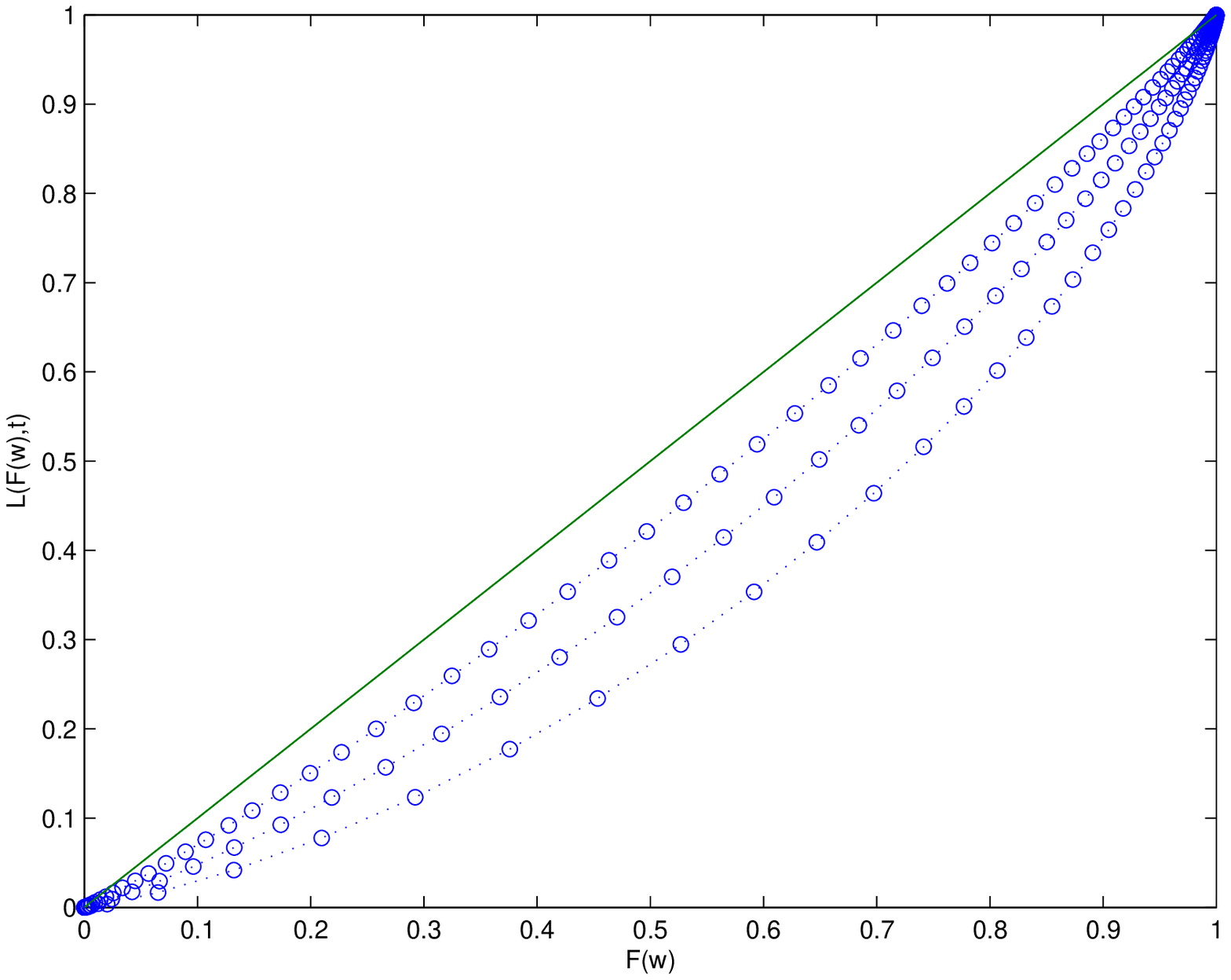}
\end{center}
\caption{Test 3. The corresponding Lorentz curves. The Gini
coefficients are $G=0.1$, $G=0.2$ and $G=0.3$ respectively.}
\label{fg:test3b}
\end{figure}

In the last test case we consider the asymptotic limit of the
Boltzmann model and compare its numerical self-similar solution
with the explicit one computed in the last section using the
Fokker-Planck model. To this end, we consider the self-similar
scaling (\ref{eq:wc}) and compute the solution for the values
$r=0.001$, $D=0.0015$ with $\xi$ and $\eta/S(0)$ distributed with
standard deviation $0.05$. We report the numerical solution for a
constant value of $\mu=0.5$ at different times $t=50,200,500$ in
Figures \ref{fg:test3} and \ref{fg:test3c}. A very good agreement
between the Boltzmann and the lognormal Fokker-Planck solutions is
observed, as expected from the results of the last section. We also
compute the corresponding Lorentz curve $L(F(w,t))$ defined as
\[
L(F(w,t))=\frac{\displaystyle\int_0^w
f(v,t)v\,dv}{\displaystyle\int_0^\infty f(v,t)v\,dv},\quad
F(w,t)=\int_0^w f(v,t)\,dv,
\]
and the Gini coefficient $G\in [0,1]$
\[
G=1-2 \int_0^1 L(F(w,t))\,dw.
\]
The Gini coefficient is a measure of the inequality in the wealth
distribution \cite{Gi}. A value of $0$ corresponds to the line of
perfect equality depicted in Figure \ref{fg:test3b} together with
the different Lorentz curves. It is clear that inequalities grow
in time due to the speculative dynamics.

\section{Conclusions}
We have derived a simple linear mesoscopic model which describes a
financial market under the assumption that the distribution of
investments is known as a function of the price. The model is able
to describe the exponential growth of the price of the stock and the
growth of the wealth above the rate produced by simple investments
in bonds. The long-time behavior of the model has been studied
with the help of a Fokker-Planck approximation. The emergence of a
power law tail for the wealth distribution of lognormal type has
been proved. In order to produce the effect of a real financial
market, with booms, cycles and crashes, the distribution of
investments should be a function of time (the decision-making
should be done by maximizing the expected utility) and one should
consider heterogeneous populations of investors as in \cite{LLS,
LLS1}. In this case, the model should be modified and the
time evolution of $\mu(S,t)$ considered. Another interesting research
direction is related to the possibility to introduce stock options into the
model and to relate the kinetic approach to
Black-Scholes type equations. All these subjects are actually
under investigation and we hope to present other challenging
results in the near future.

\subsection*{Acknowledgment}
Lorenzo Pareschi would like to thank the international research
center Le Studium and the MAPMO in Orl\'eans, France for their
support and kind hospitality during his visits. The authors also
thank J.~Carrillo and G.~Toscani for helpful discussions and David
Hunter for careful reading of the manuscript.

\end{document}